\documentclass[11pt]{article}
\usepackage{amsmath}
\usepackage{amsfonts,amssymb}
\begin{document}

\newcommand{\nnl}{\nl[6mm]}
\newcommand{\enl}{\\[6mm]}
\newcommand{\nlb}[1]{\nl[-3mm]\label{#1}\\[-3mm]}
\newcommand{\nle}{\nl[-3mm]\\[-3mm]}
\newcommand{\nl}{\nonumber\\}
\newcommand{\bl}{&&\quad}
\newcommand{\ab}{\allowbreak}

\renewcommand{\leq}{\leqslant}
\renewcommand{\geq}{\geqslant}

\renewcommand{\theequation}{\thesection.\arabic{equation}}
\let\ssection=\section
\renewcommand{\section}{\setcounter{equation}{0}\ssection}

\newcommand{\be}{\bes}
\newcommand{\ee}{\ees}
\newcommand{\bes}{\begin{eqnarray}}
\newcommand{\ees}{\end{eqnarray}}
\newcommand{\eens}{\nonumber\end{eqnarray}}

\renewcommand{\/}{\over}
\renewcommand{\d}{\partial}

\newcommand{\eps}{\epsilon}
\newcommand{\vareps}{\varepsilon}
\newcommand{\dlt}{\delta}
\newcommand{\al}{\alpha}
\newcommand{\bt}{\beta}
\newcommand{\gm}{\gamma}
\newcommand{\ka}{\kappa}
\newcommand{\si}{\sigma}
\newcommand{\la}{\lambda}
\newcommand{\rep}{\varrho}

\newcommand{\oj}{{\mathfrak g}}
\newcommand{\tr}{\hbox{tr}\,}

\newcommand{\xmu}{\xi^\mu}
\newcommand{\xnu}{\xi^\nu}
\newcommand{\xrho}{\xi^\rho}
\newcommand{\ynu}{\eta^\nu}
\newcommand{\yrho}{\eta^\rho}
\newcommand{\dmu}{\d_\mu}
\newcommand{\dnu}{\d_\nu}
\newcommand{\drho}{\d_\rho}
\newcommand{\frho}{\phi_\rho}
\newcommand{\dsi}{\d_\si}
\newcommand{\dtau}{\d_\tau}
\newcommand{\dka}{\d_\ka}
\newcommand{\dla}{\d_\la}

\newcommand{\lrst}{(\la\rho)(\si\tau)}
\newcommand{\muN}{{\mu_1..\mu_N}}
\newcommand{\sip}{{\si_1..\si_p}}
\newcommand{\tauq}{{\tau_1..\tau_q}}
\newcommand{\mn}{{m_1..m_n}}

\newcommand{\mxn}{{m \leftrightarrow n}}
\newcommand{\fxg}{{f \leftrightarrow g}}
\newcommand{\xxy}{{\xi \leftrightarrow \eta}}
\newcommand{\XxY}{{X \leftrightarrow Y}}

\newcommand{\mm}{{\mathbf m}}
\newcommand{\mtn}{{(m\times n)}}

\newcommand{\LL}{{\cal L}}
\newcommand{\HH}{{\cal H}}
\newcommand{\J}{{\cal J}}

\newcommand{\Lxi}{\LL_\xi}
\newcommand{\Leta}{\LL_\eta}
\newcommand{\LX}{\LL_\X}
\newcommand{\LY}{\LL_\Y}

\newcommand{\Grst}{{\Gamma^\rho_{\si\tau}}}

\newcommand{\into}{\hookrightarrow}
\newcommand{\e}{{\rm e}}
\newcommand{\symm}{{\rm symm}}

\newcommand{\bra}[1]{\big{\langle}#1\big{|}}
\newcommand{\ket}[1]{\big{|}#1\big{\rangle}}
\newcommand{\no}[1]{{\,:\kern-0.7mm #1\kern-1.2mm:\,}}

\newcommand{\RR}{{\mathbb R}}
\newcommand{\CC}{{\mathbb C}}
\newcommand{\ZZ}{{\mathbb Z}}

\title{{Extensions of diffeomorphism and current algebras}}

\author{T. A. Larsson \\
Vanadisv\"agen 29, S-113 23 Stockholm, Sweden,\\
 email: tal@hdd.se \\
{\tt math-ph/0002016} }

\maketitle
\begin{abstract}
Dzhumadil'daev has classified all tensor module extensions
of $diff(N)$, the diffeomorphism algebra in $N$ dimensions, and its
subalgebras of divergence free, Hamiltonian, and contact vector fields.
I review his results using explicit tensor notation.
All of his generic cocycles are limits of trivial cocycles, and many
arise from the Mickelsson-Faddeev algebra for $gl(N)$.
Then his results are extended to some non-tensor modules, including 
the higher-dimensional Virasoro algebras found by Eswara Rao/Moody 
and myself.
Extensions of current algebras with $d$-dimensional representations
are obtained by restriction from $diff(N+d)$. This gives a connection
between higher-dimensional Virasoro and Kac-Moody cocycles, and
between Mickelsson-Faddeev cocycles for diffeomorphism and current
algebras. 
\end{abstract}

\section{Introduction}
An extension $\hat L$ of a Lie algebra $L$ by a module $M$ is an exact 
sequence
\[
0 \longrightarrow M \stackrel{\imath}{\longrightarrow} \hat L 
\stackrel{\pi}{\longrightarrow} L \longrightarrow 0.
\]
This means that $\imath$ is injective, $\pi$ is surjective, and
$M$ is an ideal in $\hat L$. It is precisely this situation which is of
interest in physics, because if $L$ is a classical symmetry algebra
(realized in terms of Poisson brackets), quantum corrections are of 
order $\hbar$ and thus generate an ideal. In particular, if $M = \CC$
we say that the extension is central, which is the case that has 
attracted most attention in physics; suffice it to mention the
ample applications of Virasoro and affine Kac-Moody algebras.

The best known non-central extension is the Mickelsson-Faddeev (MF) 
algebra
\cite{Fad84,Mic85,Mic89}, which is an abelian extension of the algebra 
$map(N,\oj)$ of maps from $N$-dimensional spacetime to a 
finite-dimensional Lie algebra $\oj$. $map(N,\oj)$ also admits
higher-dimensional generalizations of the Kac-Moody cocycle
\cite{Kas85,Lar91,MEY90}, whose Fock representations were first 
constructed in \cite{EMY92,MEY90}. 
Similarly, the diffeomorphism algebra in $N$ dimensions, $diff(N)$,
has non-central extensions analogous to the Virasoro algebra
\cite{ERM94,Lar89,Lar91}. The representation theory of these algebras
was developped in \cite{BB98,Bil97,ERM94,Lar97,Lar98,Lar99}.
It appears that representations of the MF algebra, if they exist, are
not attainable by similar methods \cite{Lar00}.

In \cite{Dzhu96}, Dzhumadil'daev classified all extensions of
$L = diff(N)$ when $M$ is a tensor module. He also covered the cases when
$L$ is one of the algebras of divergence free, Hamiltonian and contact
vector fields. Unfortunately, his paper is
not easy to read for a physicist (at least not for this one), so
one purpose of the present paper is to review his classification in a
more physicist-friendly manner, using notation from tensor calculus.
Moreover, his results are quite bewildering, since he obtain no less than 
seventeen different cocycles. However, it turns out that all of them
can be grouped into four classes: 
\begin{enumerate}
\item
A cubic (in derivatives) cocycle, which splits into its traceless and
trace parts.
\item
Quartic cocycles, which follow from the MF extension for $gl(N)$
(recall that tensor fields are functions with values in $gl(N)$ modules).
In fact, only the tensor part of the MF extension was used, which
can be removed by a redefinition \`a la Cederwall et al. \cite{CFNW94}.
\item
Quintic cocycles which only exist in two dimensions.
\item
Special cases in one  dimension.
\end{enumerate}
Dzhumadil'daev also considered cocycles for the $diff(N)$ subalgebras
$svect(N)$ (divergence-free vector fields), $Ham(N)$ (Hamiltonian vector
fields) and $K(N)$ (contact vector fields). All such cocycles follow
directly by restriction from the full diffeomorphism algebra.

I extend Dzhumadil'daev's result by constructing some extensions
where $M$ is not a tensor module. These cases include the 
higher-dimensional Virasoro algebras of Eswara Rao and Moody \cite{ERM94} 
and myself \cite{Lar91}. 
Another generalization is found be considering the inhomogenous
term in the MF extension. 

It turns out that all of Dzhumadil'daev's generic cocycles and several
of his low-dimensional ones can be obtained as limits of trivial cocycles. 
One constructs
a family of trivial cocycles, parametrized by a continuous parameter
(the conformal weight $\la$), and let $\la$ approach a critical value
$\la_0$. Thus, these cocycles are non-trivial in the usual 
cohomological sense, but belong to the closure of the space of trivial 
cocycles. On the other hand, the Virasoro and MF extensions do not belong
to this closure, because there is no continuous parameter that can be
varied. Geometrically, they involve closed chains 
(one- and three-, respectively), and the closedness
condition is consistent for $\la=1$ only.

Dzhumadil'daev \cite{Dzhu92} and Ovsienko and Roger \cite{OR96} have
also classified cocycles for the special case $N=1$. I show that some of 
these have higher-dimensional analogues not previously considered,
although this generalization is quite unnatural and uninteresting.
Moreover, in one dimension these cocycles, possibly with one
exception, are limits of trivial cocycles.

Dzhumadil'daev's classification can be used to construct interesting
extensions of subalgebras of the diffeomorphism algebra. To this end,
I consider the inclusion $map(N, diff(d)) \subset diff(N+d)$,
and further $\oj \subset gl(d) \subset diff(d)$. In fact, the algebra
$map(N, diff(d))$ defines an interesting generalization of gauge symmetry:
the replacement of a global symmetry $\oj$ by a gauge symmetry
$map(N,\oj)$ amounts to a localization in base space, but the gauge
transformations are still rigid in target space. Replacing $\oj$ by 
$diff(d)$ makes transformations local in target space as well.
By studying the restriction of the extensions under
the above inclusions, existence of extensions for subalgebras is 
shown, but neither non-triviality nor exhaustion. However, non-triviality
can be checked by hand, and my method tautologically exhaust all
extensions that can be lifted to tensor module extensions of the algebra 
of diffeomorphisms in total space.

\section{Background}
\subsection{Diffeomorphism algebra}
Let $\xi=\xmu(x)\dmu$, $x\in\RR^N$, $\dmu = \d/\d x^\mu$,
be a vector field, with commutator 
$[\xi,\eta] \equiv \xmu\dmu\ynu\dnu - \ynu\dnu\xmu\dmu$.
Greek indices $\mu,\nu = 1,2,\ldots,N$ label the 
spacetime coordinates and the summation convention is used on all kinds 
of indices. The diffeomorphism algebra $diff(N)$ is generated by Lie 
derivatives $\Lxi$. Dzhumadil'daev denotes this algebra by $W_N$ in
honour of Witt. In the literature, it is also known as the algebra of
vector fields and is denoted by $Vect(N)$ or $Vect(\RR^N)$.

An extensions of $diff(N)$ is given by a bilinear cocycle $c(\xi,\eta)$:
\be
[\Lxi,\Leta] = \LL_{[\xi,\eta]} + c(\xi,\eta).
\label{ext1}
\ee
For convenience, some formulas are also displayed in a Fourier basis.
With $L_\mu(m) = \Lxi$ for $\xi = i\exp(im_\rho x^\rho)\dmu$, 
$m\in\RR^N$, (\ref{ext1}) is replaced by
\be
[L_\mu(m), L_\nu(n)] = n_\mu L_\mu(m+n) - m_\nu L_\nu(m+n) 
 + c_{\mu\nu}(m,n).
\label{ext2}
\ee
We say that an extension is {\em local} if it has the form
\be
c_{\mu\nu}(m,n) = pol^a_{\mu\nu}(m,n)A_a(m+n),
\label{local}
\ee
where $pol^a_{\mu\nu}(m,n) = -pol^a_{\nu\mu}(n,m)$ is a polynomial 
and $A_a$ is some operator.

Let $T^\mu_\nu$ form a basis for $gl(N)$, with brackets
\be
[T^\mu_\nu,T^\si_\tau] = 
\dlt^\si_\nu T^\mu_\tau - \dlt^\mu_\tau T^\si_\mu.
\label{glN}
\ee
Then
\be
\Lxi = \xmu(x)\dmu + \dnu\xmu(x)T^\nu_\mu 
\label{tensor}
\ee
satisfies (\ref{ext1}) with zero cocycle. Analogously, if $T^\mu_\nu(m)$ 
form a basis for an extension of $map(N,gl(N))$, with brackets
\bes
[T^\mu_\nu(m),T^\si_\tau(n)] &=& 
\dlt^\si_\nu T^\mu_\tau(m+n) - \dlt^\mu_\tau T^\si_\nu(m+n)
+ k^{\mu\si}_{\nu\tau}(m,n),
\nlb{glNext}
{[}L_\mu(m),T^\si_\tau(n)] &=& n_\mu T^\si_\tau(m+n),
\eens
then
\be
L'_\mu(m) = L_\mu(m) + m_\nu T^\nu_\mu(m)
\label{Tredef}
\ee
satisfies an extension of $diff(N)$ with cocycle
\be
c_{\mu\nu}(m,n) = m_\si n_\tau k^{\si\tau}_{\mu\nu}(m,n).
\label{ck}
\ee

A {\em tensor module} is the carrying space of the $diff(N)$ 
representation obtained by substituting a $gl(N)$ representation into
(\ref{tensor}). A tensor of type $(p,q;\la)$ ($p$ contravariant and
$q$ covariant indices and conformal weight $\la$)
is described by the equivalent formulas 
\bes
&&[\Lxi, \Phi^\sip_\tauq(x)]
= -\xmu(x)\dmu\Phi^\sip_\tauq(x) -
\la \dmu\xmu(x) \Phi^\sip_\tauq(x) -\nl
&&+\sum_{i=1}^p \dmu\xi^{\si_i}(x) \Phi^{\si_1..\mu..\si_p}_\tauq(x)
-\sum_{j=1}^q \d_{\tau_j}\xmu(x) \Phi^\sip_{\tau_1..\mu..\tau_q}(x), 
\nnl
&&[\Lxi, \Phi^\sip_\tauq(\phi_\sip^\tauq)]
= \Phi^\sip_\tauq(\xmu\dmu\phi_\sip^\tauq + 
(1-\la) \dmu\xmu \phi_\sip^\tauq - \nl
&&+\sum_{i=1}^p \d_{\si_i}\xmu \phi_{\si_1..\mu..\si_p}^\tauq
-\sum_{j=1}^q \dmu\xi^{\tau_j} \phi_\sip^{\tau_1..\mu..\tau_q}), 
\label{tmod} \\ \nl
&&{[}L_\mu(m), \Phi^\sip_\tauq(n)]
= (n_\mu+(1-\la)m_\mu) \Phi^\sip_\tauq(m+n) + \nl
&&+\sum_{i=1}^p \dlt^{\si_i}_\mu m_\rho 
\Phi^{\si_1..\rho..\si_p}_\tauq(m+n)
-\sum_{j=1}^q m_{\tau_j} \Phi^\sip_{\tau_1..\mu..\tau_q}(m+n),
\eens
where $\phi_\sip^\tauq$ is an arbitrary function on $\RR^N$ and
\be
\Phi^\sip_\tauq(\phi_\sip^\tauq) 
= \int d^Nx\ \phi_\sip^\tauq(x) \Phi^\sip_\tauq(x).
\ee
For brevity, we shall often write the rhs of (\ref{tmod}) simply as 
$(p,q;\la)$. Further, we abbreviate the action on objects which contain 
additional terms as
\be
{[}L_\mu(m), \Phi^\sip_\tauq(n)] = (p,q;\la) + \hbox{more}, 
\label{abbrev}
\ee
etc.
Tensor modules contain irreducible submodules consisting of symmetric,
anti-symmetric, and traceless tensors, labelled by the irreps of
$gl(N)$. As is standard in physics, (anti-)symmetrization 
of indices is denoted by parentheses (brackets), and 
vertical bars inhibit the operation. Thus, 
$\phi^{(\mu|\nu|\rho)} = \phi^{\mu\nu\rho} + \phi^{\rho\nu\mu}$
and $\phi_{[\mu\nu]} = \phi_{\mu\nu} - \phi_{\nu\mu}$.

However, not all modules are tensor modules. The following cases
will be considered below:

1. A totally skew tensor $\omega_\sip = \omega_{[\sip]}$ of type
$(0,p;0)$, i.e. a $p$-form, contains a submodule consisting of
closed $p$-forms, which amounts to the conditions
\be
\d_{[\nu}\omega_{\sip]}(x) \equiv 0, \qquad
\omega_{\sip}(\dnu\phi^{[\nu\sip]}) \equiv 0, \qquad
m_{[\nu}\omega_{\sip]}(m) \equiv 0.
\label{form}
\ee

2. Dually, a totally skew tensor $S^\sip = S^{[\sip]}$ of type $(p,0;1)$, 
can be identified as a $p$-chain. Closed $p$-chains satisfy
the conditions
\be
\d_{\si_1}S^\sip(x) \equiv 0, \qquad
S^\sip(\d_{\si_1} \phi_{\si_2..\si_p}) \equiv 0, \qquad
m_{\si_1} S^\sip(m) \equiv 0.
\label{chain}
\ee

3. The connection $\Grst$ transforms as a tensor field
of type $(1,2;0)$, apart from an additional term
\bes
[\Lxi, \Grst(x)] &=& (1,2;0) + \dsi\dtau \xi^\rho, \nl
{[}\Lxi, \Grst(\phi^{\si\tau}_\rho)] &=& 
 (1,2;0) + \int d^Nx\ \dsi\dtau \xi^\rho(x) \phi^{\si\tau}_\rho(x), 
\label{lc-conn}\\
{[}L_\mu(m),\Grst(n)] &=& (1,2;0) + m_\si m_\tau \dlt^\rho_\mu\dlt(m+n),
\eens
where $(1,2;0)$ denote regular terms as in (\ref{abbrev}).

\subsection{ Divergence-free vector fields }
The algebra of divergence-free (or special) vector fields 
$svect(N)\subset diff(N)$ is generated by $\Lxi$ such that $\dmu\xmu=0$,
or equivalently $L_\mu(m)$ such that $m_\mu = 0$. Tensor modules are
given by (\ref{tmod}), except that the conformal weight $\la$ is 
irrelevant.

\subsection{ Hamiltonian vector fields }
\label{ssec:Ham}
The algebra $Ham(N)\subset diff(N)$ ($N$ even) consists of vector fields
$\xi$ that leave the two-form $\eps_{\mu\nu}dx^\mu\wedge dx^\nu$
invariant, where $\eps_{\mu\nu}=-\eps_{\nu\mu}$ is the symplectic form, 
whose inverse $\eps^{\mu\nu}$ satisfies $\eps^{\mu\rho}\eps_{\rho\nu} 
= \eps_{\nu\rho}\eps^{\rho\mu} = \dlt^\mu_\nu$. 
Such Hamiltonian vector fields are
of the form $\xi = \eps^{\mu\nu}\dmu f\dnu$.
Dzhumadil'daev denotes $Ham(N)$ by $H_n$, where $N = 2n$.
Any extension of $Ham(N)$ takes 
the form
\be
[H_f, H_g] = H_{\{f,g\}} + c_H(f,g), 
\label{Ham}
\ee
where $\{f,g\} = \eps^{\mu\nu}\dmu f\dnu g$ is the Poisson bracket.
Alternatively, in the Fourier basis
\be
[H(m), H(n)] = \mtn H(m+n) + c_H(m,n), 
\ee
where $m\times n \equiv \eps^{\mu\nu}m_\mu n_\nu$.
Since $\eps^{\mu\nu}\dmu\dnu f \equiv 0$, any Hamiltonian vector field
is divergence free. The converse is also true in two dimensions, but
when $N\geq4$ there are divergence-free vector fields that are not
Hamiltonian.
One checks that $[H_f,\eps^{\mu\nu}] \equiv 0$, so 
$\eps^{\mu\nu}$ and $\eps_{\mu\nu}$ can be used to raise and lower
indices. A typical tensor module is hence of the form
\bes
[H_f, \Phi^\si(\phi_\si) &=& 
 \Phi^\si(\{f,\phi_\si\} + \eps^{\mu\nu}\dmu\dsi f\phi_\nu) \nle
{[}H(m), \Phi^\si(n)] &=&
 n_\mu\Phi^\si(m+n) + \eps^{\mu\si}m_\mu m_\nu \Phi^\nu(m+n).
\eens

\subsection{ Contact vector fields }
The contact algebra $K(N)\subset diff(N)$ ($N$ odd) consists of
vector fields which leave the one-form
$\al = dx^0 + \eps_{ij}x^idx^j$ invariant, where $\eps_{ij}$ is the 
symplectic form in one dimension less. Here greek indices 
$\mu,\nu=0,1,2,...,N$ run over all $N+1$ indices but latin indices
$i,j=1,2,...,N$ exclude the time index $0$. Contact vector fields are
of the form 
\be
K_f = \Delta f\d_0 + \d_0 f x^i\d_i + \eps^{ij}\d_j f\d_i,
\label{xif}
\ee
where $\Delta f = 2f - x^i\d_i f$. Any extension of $K(N+1)$ has the form
\be
[K_f, K_g] &=& K_{[f,g]_K} + c_K(f,g),
\ee
where the contact bracket reads
\be
[f,g]_K = \d_0 f\Delta g - \d_0 g\Delta f + \{f,g\}
\ee
and $\{f,g\} = \eps^{ij}\d_i f\d_j g$ is the Poisson bracket in 
$N$ dimensions.
Due to the explicit appearence of $x^i$ in the definition of $\Delta$,
it is inconvenient to describe this algebra in a Fourier basis.
Dzhumadil'daev denotes $K(N)$ by $K_n$, where $N = 2n+1$.

\subsection{ Gauge algebra }
Let $\oj$ be a finite-dimen\-sional Lie algebra with basis $J^a$ 
(hermitian if $\oj$ is compact and semisimple), 
structure constants $f^{ab}{}_c$ and brackets
$[J^a,J^b] = if^{ab}{}_c J^c$. 
Our notation is similar to \cite{GO86} or \cite{FMS96}, chapter 13.
We always assume that $\oj$ has a
Killing metric proportional to $\dlt^{ab}$. 
Further assume that there is a priviledged vector 
$\dlt^a \propto \tr J^a$, such that $f^{ab}{}_c \dlt^c \equiv 0$.
Of course, $\dlt^a = 0$ if $\oj$ is semisimple, but it may be non-zero
if $\oj$ contains abelian factors. The primary example is $\oj = gl(N)$,
where $\tr(T^\mu_\nu) \propto \dlt^\mu_\nu$.

Let $map(N,\oj)$ be the algebra of maps from $\RR^N$ to $\oj$, also
known as the gauge or current algebra. We denote its generators by 
$\J_X$, where $X=X_a(x)J^a$, $x\in\RR^N$, is a $\oj$-valued
function, and define $[X,Y] = if^{ab}{}_c X_aY_bJ^c$.
Alternatively, we use a Fourier basis with generators $\J^a(m)$,
$m\in\RR^N$. Any extension of $map(N,\oj)$ has the form
\bes
[\J_X, \J_Y] &=& \J_{[X,Y]} + c(X,Y), \nl
{[}\Lxi, \J_X] &=& \J_{\xmu\dmu X} + c(\xi,X), \nle
{[}\J^a(m), \J^b(n)] &=& if^{ab}{}_c \J^c(m+n) + c^{ab}(m,n), \nl
{[}L_\mu(m), \J^a(n)] &=& n_\mu \J^a(m+n) + c^a_\mu(m,n).
\eens
The analogue of tensor modules are functions with values in some $\oj$
module $M$. To the three formulas in (\ref{tmod}) correspond
\bes
[\J_X, \Phi^i(x)] &=& -X_a \si^{ia}_j \Phi^j(x), \nl
{[}\J_X, \Phi^i(\phi_i)] &=& -\Phi^i(X_a \si^{ja}_i\phi_j), 
\nlb{Jmod}
{[}\J^a(m), \Phi^i(n)] &=& -\si^{ia}_j \Phi^j(m+n), 
\eens
where $\si^a = (\si^{ia}_j)$ are the representation matrices acting
on $M$. Moreover, the intertwining action of diffeomorphisms is given
by (\ref{tmod}).

Among non-tensor modules we cite the connection, which is a central 
extension of the adjoint.
\bes
[\J_X, A^b_\nu(x)] &=& if^{ab}{}_c X_a(x) A^c_\nu(x) 
+ \dlt^{ab}\dnu X_a(x), \nl
{[}\J_X, A^b_\nu(\phi_b^\nu)] &=& 
 A^b_\nu(-if^{ac}{}_bX_a \phi_c^\nu) + 
 \int d^Nx\ \dlt^{ab}\dnu X_a(x)\phi_b^\nu(x), \nle
{[}\J^a(m), A^b_\nu(n)] &=& if^{ab}{}_c A^c_\nu(m+n) + m_\nu\dlt^{ab}.
\eens

The best known extension of $map(N,\oj)$, $N\geq3$, is the MF extension:
\bes
[\J^a(m), \J^b(n)] &=& if^{ab}{}_c \J^c(m+n) 
 + m_\mu n_\nu \HH^{ab\mu\nu}(m+n), \nl
{[}\J^a(m), \HH^{bc\mu\nu}(n)] &=& if^{ab}{}_d \HH^{dc\mu\nu}(m+n) +
\label{MF}\\
&&+ if^{ac}{}_d \HH^{bd\mu\nu}(m+n) 
 + d^{abc} m_\rho S_3^{\mu\nu\rho}(m+n), 
\eens
and all other brackets vanish. Here, $S_3^{\mu\nu\rho}$ is a closed
three-chain (\ref{chain}) and $d^{abc} = \tr J^{(a}J^bJ^{c)}$ 
are totally symmetric. In particular, in three dimensions we can write 
$\HH^{ab\mu\nu}(m) = \eps^{\mu\nu\rho} d^{abc}A^c_\rho(m)$, 
$S_3^{\mu\nu\rho}(m) = \eps^{\mu\nu\rho}\dlt(m)$, so $A^c_\rho(m)$
transforms as a connection. It was found in \cite{Lar00} that the
three-chain term constitutes an obstruction against the construction of
Fock modules.

\section{ Dzhumadil'daev's classification: $diff(N)$ cocycles }
Dzhumadil'daev classified extensions of $diff(N)$ by tensor 
modules \cite{Dzhu96}, and found 17 inequivalent ones. 
A tensor density can be viewed as a function with
values in a $gl(N)$ module. Since $gl(N)\sim sl(N)\oplus gl(1)$,
$gl(N)$ irreps are labelled by an $sl(N)$ highest weight and a
conformal weight $\la$. To an $sl(N)$ highest weight
$\ell_1\pi_1 + \ell_2\pi_2 + \ldots + \ell_{N-1}\pi_{N-1}$, 
$\pi_i$ being the fundamental weights, we can 
associate a partition $\{\la_1, \la_2, \ldots, \la_{N-1}\}$, where
$\la_i = \ell_i + \ell_{i+1} + \ldots + \ell_{N+1} 
= \sum_{j=i}^{N-1} \ell_j$. This can be visualized as a Young tableaux 
with $\la_i$ boxes in the $i$:th row. A contravariant vector 
corresponds to a tableaux with a single box, i.e. to the root $\pi_1$.
A covariant vector can be identified with a tableaux with a single
column with $N-1$ boxes, i.e. the root $\pi_{N-1}$.

Let $\eps_\muN$ be the totally antisymmetric symbol, given by
$\eps_\muN = +1$ if $\muN$ is a even permutation of $12..N$,
$=-1$ if it is an odd permutation, and $=0$ if two indices are equal.
It can be regarded as a constant tensor field of type $(0,N;-1)$,
since this makes the transformation law $[\Lxi,\eps_\muN(x)] = 0$
consistent. Alternatively, we may view it as a constant tensor field
$\eps^\muN$ of type $(N,0;1)$. The existence of this symbol 
establishes the standard isomorphism between $p$ lower indices and
$N-p$ upper indices, e.g. $p$-forms and $(N\!-\!p)$-chains:
\be
A_{[\mu_1..\mu_p]} = 
\eps_{\mu_1..\mu_p\nu_1..\nu_{N-p}}A^{[\nu_1..\nu_{N-p}]}.
\ee
In particular, a covariant vector field (of weight $\la$) is 
equivalent to a skew tensor field with $N-1$ contravariant indices and 
weight $\la+1$.

Dzhumadil'daev's classification is encoded in \cite{Dzhu96}, Table 1. 
The following three tables describe the corresponding modules, both
in his notation and tensor calculus notation.

1. For every $N\geq N_0$:
\be
\begin{array}{lllllll}
 & N_0 &\hbox{HW} & \la  &\hbox{Type} & \hbox{Tensor} \cr
\psi^W_1 & 1 &\pi_1 & 1 & (1,0;1) & S^\rho \cr
\psi^W_2 & 2 & 2\pi_1 + \pi_{N-1} & 2 & (2,1;1) & K^{(\si\tau)}_\mu \cr
\psi^W_{3,4} & 2& \pi_2 & 1 & (2,0;1) & F^{[\mu\nu]} \cr
\psi^W_{5,6} & 3 & \pi_1+\pi_2+\pi_{N-1} & 2
& (3,1;1) & E^{\rho(\si\tau)}_\mu : E^{(\rho\si\tau)}_\mu=0 \cr
\psi^W_7 & 2 & 3\pi_1+\pi_{N-1} & 2 
& (3,1;1) & D^{(\rho\si\tau)}_\mu \cr
\psi^W_8 & 3 & 2\pi_1+\pi_2+2\pi_{N-1} & 3 
& (4,2;1) & B^{[\lrst]}_{(\mu\nu)} \cr
\psi^W_9 & 3 & 4\pi_1+\pi_{N-2} & 2
& (4,2;1) & A^{(\la\rho\si\tau)}_{[\mu\nu]} \cr
\psi^W_{10} & 4 & 2\pi_2+\pi_{N-2} & 2
& (4,2;1) & C^{(\lrst)}_{[\mu\nu]} : 
 C^{(\la\rho\si\tau)}_{[\mu\nu]} = 0
\end{array}
\label{tabW1}
\ee
The symmetry conditions can alternatively be written as
\bes
A^{\la\rho\si\tau}_{\mu\nu} &=& A^{\rho\la\si\tau}_{\mu\nu}
= A^{\si\rho\la\tau}_{\mu\nu} = A^{\tau\rho\si\la}_{\mu\nu}
= -A^{\la\rho\si\tau}_{\nu\mu}, \nl
B^{\la\rho\si\tau}_{\mu\nu} &=& B^{\rho\la\si\tau}_{\mu\nu}
= -B^{\si\tau\la\rho}_{\mu\nu}
= B^{\la\rho\si\tau}_{\nu\mu}, \nl
C^{\la\rho\si\tau}_{\mu\nu} &=& C^{\rho\la\si\tau}_{\mu\nu}
= -C^{\si\rho\la\tau}_{\mu\nu} = C^{\si\tau\la\rho}_{\mu\nu}
= -C^{\la\rho\si\tau}_{\nu\mu}, \nle
D^{\rho\si\tau} &=& D^{\rho\tau\si} = D^{\si\rho\tau}, \nl
E^{\rho\si\tau} &=& E^{\rho\tau\si} = -E^{\si\rho\tau}, \nl
F^{\mu\nu} &=& - F^{\nu\mu}, \nl
K^{\si\tau}_\mu &=& K^{\tau\si}_\mu,
\eens
in addition to total tracelessness.

2. In addition for $N=2$:
\be
\begin{array}{llllll}
\hbox{Name} &\hbox{HW} & \la & \hbox{Partition} &
\hbox{Type} & \hbox{Tensor} \cr
\psi^W_{11},\psi^W_{12} &\pi_1 & 0 & \{1\} & (1,0;0) & S^\rho \cr
\psi^W_{13} &5\pi_1 & 2 & \{5\} & (5,0;2) & S^{(\rho_1..\rho_5)} \cr
\psi^W_{14} &7\pi_1 & 3 & \{7\} & (7,0;3) & S^{(\rho_1..\rho_7)}
\end{array}
\label{tabW2}
\ee

3. In addition for $N=1$, there are three cocycles with $\la = -1$, 
$-4$, and $-6$, respectively.

Of course, knowledge of the relevant module is not enough to uniquely 
describe the cocycle. Dzhumadil'daev has also given formulas for
the cocycles, but
it is in fact quite easy to reconstruct them from scratch. By writing
down manifestly non-trivial cocycles valued in the right modules, we
are guaranteed to obtain expressions that are equivalent to
Dzhumadil'daev's, without having to decipher his notation. This is
the subject of the rest of this section.

\subsection{$\psi^W_1$}
$\psi^W_1$ corresponds to the partition $\{1,0,\ldots,0\}$, i.e. a
tensor density $S^\rho$ of type $(1,0;1)$.
This extension was first described in \cite{Lar89}:
\bes
c(\xi,\eta) &=& S^\rho(\drho\dmu\xmu\dnu\ynu - \dmu\xmu\drho\dnu\ynu), 
\nlb{Sext}
c_{\mu\nu}(m,n) &=& m_\mu n_\nu (m_\rho-n_\rho)S^\rho(m+n).
\eens
Such a tensor can be identified with a one-chain, which is reducible
according to (\ref{chain}). It was noted in \cite{Lar91} that
(\ref{Sext}) still defines a cocycle when restricted to the submodule
of closed one-chains.
In one dimension, $\psi^W_1$ is related to the Virasoro algebra. We
have
\bes
[L_m, L_n] &=& (n-m)L_{m+n} + (n-m)mnS_{m+n},
\nlb{Svir}
[L_m, S_n] &=& (n+m)S_{m+n}.
\eens
The closedness condition, $mS_m = 0$, has the unique solution
$S_m = c/24 \dlt_m$. Substituting this into (\ref{Svir}) yields the
Virasoro algebra with central charge $c$.

Let $X$ be a function on $\RR^N$ and $E_X$ a tensor density of type 
$(0,0;1)$. If
\be
[\Lxi, E_X] = E_{\xi X}, \qquad 
[E_X, E_Y] = S^\rho(X\drho Y-\drho XY),
\ee
then $\Lxi' = \Lxi + E_{\dmu\xmu}$ satisfies $diff(N)$ with cocycle 
$\psi^W_1$. 

\subsection{$\psi^W_2$}
$\psi^W_2$ corresponds to the partition $\{3,1,\ldots,1\}$, i.e.
a traceless tensor density $K^{(\si\tau)}_\mu$ of type $(2,1;1)$:
\bes
c(\xi,\eta) &=& K^{(\si\tau)}_\nu(\dmu\xmu\dsi\dtau\ynu)
 - K^{(\si\tau)}_\mu(\dsi\dtau\xmu\dnu\ynu), 
\nlb{Kext}
c_{\mu\nu}(m,n) &=& m_\mu n_\si n_\tau K^{(\si\tau)}_\nu(m+n)
 - n_\nu m_\si m_\tau K^{(\si\tau)}_\mu(m+n).
\eens
Change the weight to some $\la\neq1$, i.e. $K^{(\si\tau)}_\nu$ is of
type $(2,1;\la)$, and redefine the $diff(N)$ generators by
\be
L_\mu(m) \mapsto L_\mu(m) + a m_\si m_\tau K^{(\si\tau)}_\mu(m).
\label{Kredef}
\ee
The new generators satisfy $diff(N)$ with the extension
$a(1-\la)c_{\mu\nu}(m,n)$, which thus is trivial. 
If we now fix $a = (1-\la)^{-1}$ and let $\la\to 1$, $\psi^W_2$
is recovered. Tracelessness does not play a role here; setting
$K^{(\si\tau)}_\mu = \dlt^{(\si}_\mu S^{\tau)}$, we see that the
trace is of type $\psi^W_1$.

\subsection{$\psi^W_3$--$\psi^W_{10}$}
These eight extensions all follow from the following
reducible extension
\bes
c(\xi,\eta) &=& 
R^{\lrst}_{\mu\nu}(\dla\drho\xmu\dsi\dtau\ynu),
\nlb{Rext}
c_{\mu\nu}(m,n) &=& m_\la m_\rho n_\si n_\tau
 R^{\lrst}_{\mu\nu}(m+n),
\eens
where $R^{\lrst}_{\mu\nu}$ is a tensor of type
$(4,2;1)$, and
\be
R^{(\si\tau)(\la\rho)}_{\nu\mu} = -R^{\lrst}_{\mu\nu}.
\label{Rsymm}
\ee
Such a tensor can be decomposed into irreducible submodules as follows.
\bes
R^{\lrst}_{\mu\nu} &=&
A^{(\la\rho\si\tau)}_{[\mu\nu]} +
B^{[\lrst]}_{(\mu\nu)} +
C^{(\lrst)}_{[\mu\nu]} + 
\label{Rdec}\\
&&\dlt^{(\la}_\mu G^{\rho)(\si\tau)}_\nu - 
\dlt^{(\si}_\nu G^{\tau)(\la\rho)}_\mu + 
\dlt^{(\la}_\nu G^{\rho)(\si\tau)}_\mu - 
\dlt^{(\si}_\mu G^{\tau)(\la\rho)}_\nu,
\eens
where 
\bes
&&G^{\rho(\si\tau)}_\nu = D^{(\rho\si\tau)}_\nu + E^{\rho(\si\tau)}_\nu
+ \dlt^{(\si}_\nu F^{\tau)\rho}
+ \dlt^{(\si}_\nu H^{\tau)\rho}
+ \dlt^\rho_\nu H^{\si\tau}, \nl
&&C^{(\la\rho\si\tau)}_{[\mu\nu]} \equiv 0, \qquad
E^{(\rho\si\tau)}_\nu \equiv 0,
\label{GDE} \\
&&F^{\rho\si} = F^{[\rho\si]}, \qquad
H^{\rho\si} = H^{(\rho\si)}.
\eens
Substitution of (\ref{Rdec})--(\ref{GDE}) into (\ref{Rext}) yields 
\[
\begin{array}{ll}
\hbox{Cocycle}&c(\xi,\eta)\cr
\hbox{Partition} & c_{\mu\nu}(m,n) \cr\cr
\psi^W_3 &
F^{[\rho\tau]}(\drho\dmu\xmu\dtau\dnu\ynu) \cr
\{1,1,0,\ldots,0\} & 
m_\rho m_\mu n_\tau n_\nu F^{[\rho\tau]}(m+n) \cr\cr
\psi^W_4 & 
F^{[\rho\tau]}(\drho\dnu\xmu\dtau\dmu\ynu) \cr
\{1,1,0,\ldots,0\} & 
m_\rho m_\nu n_\tau n_\mu F^{[\rho\tau]}(m+n) \cr\cr
\psi^W_5 & 
E^{\rho(\si\tau)}_\nu(\drho\dmu\xmu\dsi\dtau\ynu) - \xxy \cr
\{3,2,1,\ldots,1\} &
m_\rho m_\mu n_\si n_\tau E^{\rho(\si\tau)}_\nu(m+n)-\mxn \cr\cr
\psi^W_6 &
E^{\rho(\si\tau)}_\mu(\drho\dnu\xmu\dsi\dtau\ynu) - \xxy \cr
\{3,2,1,\ldots,1\} &
m_\rho m_\nu n_\si n_\tau E^{\rho(\si\tau)}_\mu(m+n)-\mxn \cr\cr
\psi^W_7 &
D^{(\rho\si\tau)}_\nu(\drho\dmu\xmu\dsi\dtau\ynu) - \xxy \cr
\{4,1,\ldots,1\} &
m_\rho m_\mu n_\si n_\tau D^{(\rho\si\tau)}_\nu(m+n) -\mxn\cr\cr
\psi^W_8 &
B^{[\lrst]}_{(\mu\nu)}(\dla\drho\xmu\dsi\dtau\ynu) \cr
\{5,3,2,\ldots,2\} &
m_\la m_\rho n_\si n_\tau B^{[\lrst]}_{(\mu\nu)}(m+n) \cr\cr
\psi^W_9 &
A^{(\la\rho\si\tau)}_{[\mu\nu]}(\dla\drho\xmu\dsi\dtau\ynu) \cr
\{5,1,\ldots,1,0\} &
m_\la m_\rho n_\si n_\tau A^{(\la\rho\si\tau)}_{[\mu\nu]}(m+n) \cr\cr
\psi^W_{10} &
C^{(\lrst)}_{[\mu\nu]}(\dla\drho\xmu\dsi\dtau\ynu) \cr
\{3,3,1,\ldots,1,0\} &
m_\la m_\rho n_\si n_\tau C^{(\lrst)}_{[\mu\nu]}(m+n)
\end{array}
\]

Splitting $R^{\lrst}_{\mu\nu}$ as in (\ref{Rdec}) and
(\ref{GDE}) suggests that there should be additional cocycles
\bes
&&m_\rho m_\nu n_\si n_\tau D^{(\rho\si\tau)}_\mu(m+n)-\mxn, \nle
&&m_\mu m_\nu n_\si n_\tau H^{(\si\tau)}(m+n) - \mxn,
\eens
but these cocycles can be removed by the redefinitions
\bes
L_\mu(m) &\mapsto& L_\mu(m)
+ m_\rho m_\si m_\tau D^{(\rho\si\tau)}_\mu(m), 
\nlb{3redef}
L_\mu(m) &\mapsto& L_\mu(m) + m_\mu m_\si m_\tau H^{(\si\tau)}(m),
\eens
respectively.

The extension (\ref{Rext}) can be understood as a MF term (\ref{MF})
for $gl(N)$ with $S^{\mu\nu\rho}_3 = 0$. 
This MF algebra is the extension of $map(N,gl(N))$ (\ref{glNext}) with
\be
k^{\mu\si}_{\nu\tau}(m,n) =
m_\rho n_\la R^{\mu\rho\si\la}_{\nu\tau}(m+n),
\ee
where $R^{\mu\rho\si\la}_{\tau\nu}$ transforms as a tensor of type
$(4,2)$ under $gl(N)$. (\ref{Rext}) now follows immediately 
from (\ref{ck}).
Moreover, since $m_\mu m_\rho \propto m_{(\mu}m_{\rho)}$, only the part 
with symmetries (\ref{Rsymm}) is relevant. 

There is another way to arrive at the extension (\ref{Rext}). An 
analogous construction was carried out by \cite{CFNW94} in the case
of arbitrary gauge algebras $map(N,\oj)$.
Let $P^{\nu\rho}_\mu$ be a tensor field of type $(2,1;1)$.
Then
\be
\Lxi' = \Lxi + P^{\nu\rho}_\mu(\dnu\drho\xmu), \qquad
L_\mu'(m) = L_\mu(m) + m_\nu m_\rho P^{\nu\rho}_\mu(m),
\label{addP}
\ee
satisfies an extension of $diff(N)$ given by
\bes
c(\xi,\eta) &=& [P^{\la\rho}_\mu(\dla\drho\xmu), 
 P^{\si\tau}_\nu(\dsi\dtau\ynu)], \nle
c_{\mu\nu}(m,n) &=& m_\la m_\rho n_\si n_\tau 
 [P^{\la\rho}_\mu(m), P^{\si\tau}_\nu(n)].
\eens
This is of the form (\ref{Rext}), if we impose the condition that
the extension be local in the sense of (\ref{local}). 
In particular, the symmetry condition (\ref{Rsymm}) holds automatically.

\subsection{ $N=2$: $\psi^W_{11}$ -- $\psi^W_{14}$}
In two dimensions, the symplectic form $\eps_{\mu\nu}$ coincides with
the anti-symmetric symbol, and hence it commutes with all vector fields, 
not just the Hamiltonian ones. This makes it possible to construct
further cocycles:
\[
\begin{array}{ll}
\hbox{Cocycle}&c(\xi,\eta)\cr
 & c_{\mu\nu}(m,n) \cr\cr
\psi^W_{11} & S^\rho(\drho(\eps^{\si\tau}\dsi\dnu\xmu\dtau\dmu\ynu
- \eps_{\mu\nu}\eps^{\si\tau}\eps^{\rho\la}\dsi\drho\xmu\dtau\dla\ynu))\cr 
 & (m_\nu n_\mu \mtn - \eps_{\mu\nu}\mtn^2)
 (m_\rho+n_\rho)S^\rho(m+n) \cr\cr
\psi^W_{12} & S^\rho(\drho(\eps^{\si\tau}\dsi\dmu\xmu\dtau\dnu\ynu)) \cr
& m_\mu n_\nu \mtn(m_\rho+n_\rho)S^\rho(m+n) \cr\cr
\psi^W_{13} & S^{(\ka\la\rho\si\tau)}(3\eps_{\mu\ka}\eps_{\nu\la}
\eps^{\al\bt}\d_\al\drho\dsi\xmu\d_\bt\dtau\ynu
- 4 \eps_{\mu\nu} \dka\drho\dsi\xmu\dla\dtau\ynu) \cr
&\qquad- \xxy \cr
 & (3\mtn\eps_{\mu\ka}\eps_{\nu\la}m_\rho m_\si n_\tau
- 4 \eps_{\mu\nu} m_\ka m_\rho m_\si n_\la n_\tau) \cdot\cr
&\qquad\cdot S^{(\ka\la\rho\si\tau)}(m+n) - \mxn \cr\cr
\psi^W_{14} & S^{(\al\bt\ka\la\rho\si\tau)}(
\eps_{\mu\al}\eps_{\nu\bt} \dka\drho\dsi\xmu\dla\dtau\ynu)  - \xxy \cr
& \eps_{\mu\al}\eps_{\nu\bt} m_\ka m_\rho m_\si n_\la n_\tau
 S^{(\al\bt\ka\la\rho\si\tau)}- \mxn 
\end{array}
\]
As in subsection \ref{ssec:Ham}, 
$\eps^{12} = - \eps^{21} = -\eps_{12} = \eps_{21} = 1$ and 
$m\times n = m_1n_2 - m_2n_1 = \eps^{\mu\nu}m_\mu n_\nu$.

The Jacobi identities, in the Fourier basis, were verified numerically 
on a computer. To this end, it was useful to write the cocycles as
\bes
c(\xi,\eta) &=& R(\xi,\xi,\xi,\eta,\eta) - R(\eta,\eta,\eta,\xi,\xi), \\
c_{\mu\nu}(m,n) &=& R_{\mu\nu}(m,m,m,n,n|m+n) - R_{\nu\mu}(n,n,n,m,m|m+n),
\eens
where e.g.
\be
R(m,r,s,n,t|u) = m_\ka r_\la s_\rho n_\si t_\tau
R^{(\ka\la\rho)(\si\tau)}_{\mu\nu}(u),
\ee
and this operator carries conformal weight $1$. 
The Jacobi identities now lead to the conditions
\bes
n_\mu R_{\nu\si}(m,m,n,s,s|m+n+s) 
 + s_\nu R_{\si\mu}(n,n,s,m,m|m+n+s)&&\nl
+ m_\si R_{\mu\nu}(s,s,m,n,n|m+n+s)
 - s_\mu R_{\si\nu}(m,m,s,n,n|m+n+s)&&\\
- m_\nu R_{\mu\si}(n,n,m,s,s|m+n+s) 
 - n_\si R_{\nu\mu}(s,s,n,m,m|m+n+s) &=&0.
\eens
Note how the epsilons
conspire to yield the correct assignments of conformal weights: 
$\eps^{\mu\nu}$ and $\eps_{\mu\nu}$ carry weight $+1$ and $-1$,
respectively. In particular, $S^\rho$ has weight zero and is therefore
not a one-chain, so it is not possible to write 
$S^\rho(\drho F) = \Phi(F)$ for $\Phi$ a tensor density.

\subsection{ $N=1$ }
In one dimension vectors have only one component, so we can use the
simplified notation $L_m = L_1(m)$. 
A density with weight $\la$
(often called a primary field) transforms as 
$[L_m, A_n] = (n+(1-\la)m)A_{m+n}$. 
Dzhumadil'daev describes the cocycles as follows.
\be
\begin{array}{ll}
\la & \hbox{Cocycle} \cr
-1 & L_0\wedge L_2\mapsto 1 \cr
-4 & L_2\wedge L_3\mapsto 1 \cr
-6 & L_2\wedge L_5\mapsto 1, L_3\wedge L_4\mapsto -3
\end{array}
\label{tabW3}
\ee
It is not easy to guess the explicit forms of the cocycles from this
description, but fortunately they were given in \cite{Dzhu92}. 
This list was later rediscovered by
Ovsienko and Roger \cite{OR96}, and the super generalization has 
recently be worked out by Marcel \cite{Mar99}. I follow his naming
scheme for the cocycles.
\be
\begin{array}{lcl}
&\quad\la\quad & c(m,n) \cr
\gamma_1 & 1 & (m-n)A_{m+n} \cr
\gamma_2=\psi^W_1 & 0 & (m^2n-mn^2)A_{m+n} \cr
\gamma_3 & 0 & (m^2 - n^2)A_{m+n} \cr
\gamma_4 & -1 & (m^3n-mn^3)A_{m+n} \cr
\gamma_5 & -1 & (m^3 - n^3)A_{m+n} \cr
\gamma_6 & -4 & (m^3n^4-n^3m^4)A_{m+n} \cr
\gamma_7 & -6 & (2m^3n^6-9m^4n^5 + 9n^4m^5 - 2n^3m^6)A_{m+n}
\end{array}
\label{ORext}
\ee
For some reason, $\gamma_1$, $\gamma_3$ and $\gamma_5$ are not included
in Dzhumadil'daev's 1996 classification, although they are present in
his 1992 paper.
Moreover, we have the Virasoro cocycle with values in the trivial module.

Generalize primary fields to translated primary fields:
\be
[L_m,A_n] = (n+r+(1-\la) m) A_{m+n}.
\label{trans}
\ee
Now consider the redefinition
\be
L_m \mapsto L'_m = L_m + a m^p A_m,
\label{Lredef}
\ee
where $a$ is a parameter.
This redefinition gives rise to a trivial cocycle, except when 
$\la=\la_0$ and $r=0$, where $\la_0$ is the weight in the table above.
In this critical case, (\ref{Lredef}) gives rise to no cocycle at all.
Now set $a = 1/(\la-\la_0)$, $r=0$, and take the limit
$\la\to\la_0$. This limiting procedure yields the cocycle
$c_\la(m,n)$. Or set $a = 1/r$, $\la=\la_0$, and take the limit
$r\to0$, giving cocycle $c_r(m,n)$. The result is
\[
\begin{array}{cccc}
p & \quad\la_0\quad & c_\la(m,n) & (m,n) \cr
0 & 1 & \gamma_1 & - \cr
1 & any & - & \gamma_1 \cr
2 & 0 & \gamma_2 & \gamma_3 \cr
3 & -1 & \gamma_4 & \gamma_5 
\end{array}
\]
In this way, the cocycles $\gamma_1-\gamma_5$ arise as limits of
trivial cocycles.
To ``explain'' $\gamma_6$, assume that in the $p=3$ case,
\[
[A_m,A_n] = (m-n)B_{m+n},
\]
(``locality'') implying that $B$ transforms with $\la=-4$. 
The same assumption can be made also when $p=0,1,2$, but this gives 
nothing new.

\section{ Dzhumadil'daev's classification: subalgebra cocycles }

\subsection{ Divergence-free vector fields }
The cocycles are obtained by restriction from $diff(N)$.
Since $\dmu\xmu=0$, cocycles $\psi^W_1$, $\psi^W_2$, $\psi^W_3$, 
$\psi^W_6$ and $\psi^W_7$ vanish, and the remaining extensions
for $N\geq3$ are denoted in \cite{Dzhu96}, Table 2, by
\[
\begin{array}{cccccc}
svect(N) & \psi^S_1 & \psi^S_2 & \psi^S_3 & \psi^S_4 & \psi^S_5 \cr\cr
diff(N) & \psi^W_4 & \psi^W_5 & \psi^W_8 & \psi^W_9 & \psi^W_{10}
\end{array}
\]
The treatment of the special two-dimensional cocycles is deferred to
the next subsection, because $svect(2)\sim Ham(2)$

\subsection{ Hamiltonian vector fields }
$Ham(N)$ has nontrivial cocycles in the following modules:
\[
\begin{array}{llllll}
\hbox{Name} &N&\hbox{HW} & \hbox{Tensor} \cr
\psi^H_1 \hbox{(Moyal)} &N\geq2&0 & \Phi \cr
\psi^H_2 &N\geq4&\pi_2& A^{[\mu\nu]} \cr
\psi^H_3 &N\geq4&2\pi_2
 & B^{[(\mu\rho)(\nu\si)]} = \tilde B^{([\mu\nu][\rho\si])} \cr
\psi^H_4 &N\geq4&3\pi_2
 & C^{[(\mu\rho\la)(\nu\si\tau)]} 
 = \tilde C^{([\mu\nu][\rho\si][\la\tau])} \cr
\psi^H_5 &N\geq4&4\pi_1+\pi_2
 & D^{(\rho\la\si\tau)[\mu\nu]} \cr
\psi^H_6 &N\geq2&\pi_1 & S^\rho \cr
\psi^H_7 &N=2&7\pi_1 & S^{(\mu\nu\ka\la\rho\si\tau)} \cr
\psi^H_8 &N=2&2\pi_1 & S^{(\mu\nu)}
\end{array}
\]
The tensors are demanded to be totally traceless, in the following 
sense.
\be
\eps_{\mu\nu}A^{[\mu\nu]} = \eps_{\mu\nu}B^{[(\mu\rho)(\nu\si)]}
= \eps_{\mu\nu} C^{[(\mu\rho\la)(\nu\si\tau)]}
= \eps_{\mu\nu} D^{(\rho\la\si\tau)[\mu\nu]} = 0.
\ee

Explicitly, the cocycles are given by (\cite{Dzhu96}, Table 4)
\[
\begin{array}{ll}
\hbox{Cocycle}&c_H(f,g)\cr
N & c_H(m,n) \cr\cr
\psi^H_1 & \Phi(\eps^{\mu\nu}\eps^{\rho\si}\eps^{\la\tau}
\dmu\drho\dla f\dnu\dsi\dtau g)\cr
N\geq2& \mtn^3\Phi(m+n) \cr\cr
\psi^H_2 & A^{[\mu\nu]}(\eps^{\rho\si}\eps^{\la\tau}
\dmu\drho\dla f\dnu\dsi\dtau g)\cr
N\geq2& \mtn^2m_\mu n_\nu A^{[\mu\nu]}(m+n) \cr\cr
\psi^H_3 & B^{[(\mu\rho)(\nu\si)]}(\eps^{\la\tau}
\dmu\drho\dla f\dnu\dsi\dtau g)\cr
N\geq2& \mtn m_\mu m_\rho n_\nu n_\si B^{[(\mu\rho)(\nu\si)]}(m+n) \cr\cr
\psi^H_4 & C^{[(\mu\rho\la)(\nu\si\tau)]}(
\dmu\drho\dla f\dnu\dsi\dtau g)\cr
N\geq2& m_\mu m_\rho m_\la n_\nu n_\si n_\tau
 C^{[(\mu\rho\la)(\nu\si\tau)]}(m+n) \cr\cr
\psi^H_5 & D^{(\rho\la\si\tau)[\mu\nu]}(
\dmu\drho\dla f\dnu\dsi\dtau g)\cr
N\geq2& m_\mu m_\rho m_\la n_\nu n_\si n_\tau
 D^{(\rho\la\si\tau)[\mu\nu]}(m+n) \cr\cr
\psi^H_6 & S^\rho(\eps^{\mu\nu}\eps^{\ka\si}\eps^{\la\tau}
\drho(\dmu\dka\dla f\dnu\dsi\dtau g))\cr
N\geq2& \mtn^3(m_\rho+n_\rho) S^\rho(m+n) \cr\cr
\psi^H_7 & S^{\mu\nu\ka\la\rho\si\tau}(
\dmu\dka\drho\dsi f \dnu\dla\dtau g)  - \fxg \cr
N=2& m_\mu m_\ka m_\rho m_\si n_\nu n_\la n_\tau
 S^{\mu\nu\ka\la\rho\si\tau}(m+n)- \mxn, \cr\cr
\psi^H_8 & \eps^{\ka\la}\eps^{\rho\pi}\eps^{\si\tau}
 S^{\mu\nu}( 7\dmu\dka\drho\dsi f \dnu\dla\d_\pi\dtau g +\cr
&\qquad + 3 (\dmu\dnu\dka\drho\dsi f \dla\d_\pi\dtau g +
 \dka\drho\dsi f \dmu\dnu\dla\d_\pi\dtau g)) \cr
N=2&(7m_\mu n_\nu + 3(m_\mu m_\nu + n_\mu n_\nu)) \mtn^3S^{\mu\nu}(m+n)
\end{array}
\]
$\psi^H_1-\psi^H_5$ arise from restriction of the traceless cocycles
of $\psi^W_3-\psi^W_{10}$, i.e. $\psi^S_1-\psi^S_5$. The modules
appear different because we can use the symplectic form to eliminate
all lower indices. Explicitly, we have
\bes
R^{(\la\rho)(\si\tau)}_{\mu\nu} &=&
\eps_{\mu\al}\eps_{\nu\bt} R^{\al(\la\rho)\bt(\si\tau)}, 
\nlb{Rup}
R^{\bt(\si\tau)\al(\la\rho)} &=& -R^{\al(\la\rho)\bt(\si\tau)}.
\eens
In this way, each traceless module in
$\psi^W_3-\psi^W_{10}$ can be expanded as a direct sum of the modules
in $\psi^H_1-\psi^H_5$.
E.g., the field in $\psi^W_4$ can be written as
$F^{[\mu\nu]} = \eps^{\mu\nu}\Phi + A^{[\mu\nu]}$ with 
$\eps_{\mu\nu}A^{[\mu\nu]} = 0$, and thus we obtain $\psi^H_1$ and
$\psi^H_2$. The conformal weight is irrelevant for Hamiltonian vector
fields, wherefore we can consistently substitute
$\Phi(\phi) \equiv S^\rho(\drho\phi)$ in $\psi^H_1$; this gives 
$\psi^H_6$. $\psi^H_6$ also follows by restriction from $\psi^W_{11}$ in
two dimensions, but exists in all dimensions. For the special 
two-dimensional cocycles, $\psi^H_7$ is the restriction of 
$\psi^W_{14}$, whereas $\psi^W_{12}$ is a divergence which restricts to
zero. Finally, it was checked numerically that $\psi^H_8$ is a cocycle.
It may be related to $\psi^W_{13}$.

As is well known, the Moyal cocycle $\psi^W_1$ can be integrated to a
full-fledged deformation of $Ham(N)$. Consider the Moyal (or sine) 
algebra, which is the Lie algebra with brackets (in Fourier basis)
\be
[H(m), H(n)] = {1\/\hbar} sin(\hbar \mtn) H(m+n).
\ee
The Moyal cocycle appears at the lowest non-trivial order in $\hbar$.
\bes
[H(m), H(n)] &=& \mtn H(m+n) + \mtn^3\Phi(m+n), \nl
{[}H(m), \Phi(n)] &=& \mtn \Phi(m+n) \\
{[}\Phi(m), \Phi(n)] &=& 0,
\eens
where $\Phi(m) = (\hbar^2/6) H(m)$. 

\subsection{ Contact vector fields}
Dzhumadil'daev lists the $K(N)$ cocycles in his Table 5. His results
for the relevant modules and thee cocycles are
\[
\begin{array}{lllll}
\hbox{Cocycle}&HW&N&\qquad&c_K\cr
\psi^K_1 &0& N\geq 3 && \hat\psi^H_1 \cr
\psi^K_2 &\pi_2& N\geq 5 && \hat\psi^H_2 \cr
\psi^K_3 &\pi_2& N\geq 5 && \psi^W_4 \cr
\psi^K_4 &2\pi_2& N\geq 5 && \hat\psi^H_3 \cr
\psi^K_5 &3\pi_2& N\geq 5 && \hat\psi^H_4 \cr
\psi^K_6 &4\pi_1+\pi_2& N\geq 5 &&\hat\psi^H_5\cr
\psi^K_7 &2\pi_1+\pi_2& N\geq 5 \cr
\psi^K_8 &4\pi_1& N\geq 3 \cr
\psi^K_9 &\pi_1& N=3 &&  \hat\psi^H_6 = \psi^W_{11} \cr
\psi^K_{10} &\pi_1& N=3 && \psi^W_{12}\cr
\psi^K_{11} &3\pi_1& N=3 & \cr
\psi^K_{12} &5\pi_1& N=3 && \psi^W_{13}\cr
\psi^K_{13} &7\pi_1& N=3 && \psi^W_{14}
\end{array}
\]
$\psi^K_7$ and $\psi^K_8$ arise by restriction from $\psi^W_6-
\psi^W_7$.
An important point is that contact vector fields have non-zero
divergence, $\hbox{div} K_f \propto \d_0 f$.
Here $\hat\psi^H_n$ means that the restriction to the Hamiltonian 
subalgebra is $\psi^H_n$, and $\psi^W_n$ that the cocycle is obtained
by restriction from to unrestricted diffeomorphism algebra.
$\psi^K_{11}$ is a special cocycle which seems to be unique to the
contact algebra; I have not verified its existence.

However, I fail to understand Dzhumadil'daev's results on two 
points. 
\begin{enumerate}
\item
In view of the results in the previous subsection, the eight cocycles
$\psi^K_1-\psi^K_8$ arise by restriction from the eight MF cocycles
$\psi^W_3-\psi^W_{10}$. However, moving all indices upstairs as in
(\ref{Rup}) requires the existence of an invariant and invertible
two-form $\eps_{\mu\nu}$. This exists for the Hamiltonian algebra but
not, as far as I understand, for the contact algebra.
\item
$K(N)$ contains a $diff(1)$ subalgebra obtained by requiring that
the function $f(x^0)$ depends on $x^0$ only. In this case, (\ref{xif})
becomes
\be
K_f = 2f(x^0)\d_0 + f'(x^0) x^i\d_i.
\ee
These operators generate $diff(1)$ and $K_f$ is recognized as the 
expression for a primary field. Upon the restriction $diff(N)\to
K(N)\to diff(1)$, the cocycle $\psi^W_1$ for $diff(N)$ becomes
$\psi^W_1$ for $diff(1)$. Hence there must exist a nontrivial $K(N)$
cocycle with coefficients in $\pi_1$, also for $N\geq5$.
\end{enumerate}
Nevertheless, almost all cocycles follow by restriction from
$diff(N)$.

\section{ Beyond tensor modules }

\subsection{ Higher-dimensional Virasoro algebras }
We start from the tensor extensions $\psi^W_3$ and $\psi^W_4$, 
which involve the skew tensor field $F^{\rho\si}$
of type $(2,0;1)$, i.e. a two-chain. However, the extensions have the
form
\be
\begin{array}{ccc}
c(\xi,\eta) &&c_{\mu\nu}(m,n) \cr
F^{\rho\tau}(\dtau(\dmu\xmu\drho\dnu\ynu)), &\qquad&
m_\mu n_\rho n_\nu (m_\tau+n_\tau) F^{\rho\tau}(m+n), \cr
F^{\rho\tau}(\dtau(\dnu\xmu\drho\dmu\ynu)), &\qquad&
m_\nu n_\rho n_\mu (m_\tau+n_\tau) F^{\rho\tau}(m+n), 
\end{array}
\ee
respectively. By (\ref{chain}), we can now rewrite the extensions as
\be
\begin{array}{ccc}
c(\xi,\eta) &&c_{\mu\nu}(m,n) \cr
S^\rho(\dmu\xmu\drho\dnu\ynu), &\qquad&
m_\mu n_\rho n_\nu S^\rho(m+n), \cr
S^\rho(\dnu\xmu\drho\dmu\ynu), &\qquad&
m_\nu n_\rho n_\mu S^\rho(m+n),
\end{array}
\label{1ext}
\ee
where $S^\rho$ is the exact one-chain defined by
\be
S^\rho(\frho) = F^{\rho\tau}(\dtau\frho), \qquad
S^\rho(m) = m_\tau F^{\rho\tau}(m).
\ee
In particular, exact one-chains are also closed, and it turns out that
this is enough to satisfy the cocycle condition. Thus,
(\ref{1ext}) defines cocycles, to be denoted by $\bar\psi^W_3$ and
$\bar\psi^W_4$, provided that 
\be
S^\rho(\drho\phi) \equiv 0, \qquad
m_\rho S^\rho(m) \equiv 0
\ee
holds identically. $\bar\psi^W_4$ is the Eswara Rao-Moody cocycle 
\cite{ERM94}, and $\bar\psi^W_3$ was first described by myself 
\cite{Lar91}. Contrary to $\psi^W_3$ and $\psi^W_4$, these
cocycles are defined for all $N$ including $N=1$, and
in one dimension both reduce to the Virasoro cocycle.

There was some confusion in \cite{Lar97} regarding these cocycles.
The reason that they are not included in Dzhumadil'daev's list is 
that they do not involve tensor modules, but rather submodules thereof.
$\bar\psi^W_3$ and $\bar\psi^W_4$ arise naturally in toroidal Lie 
algebras.

\subsection{ Mickelsson-Faddeev }
Let $d^{\rho\tau\bt}_{\ka\nu\gm}$ be totally symmetric under interchange 
of the pairs $(\rho,\ka)$, $(\tau,\nu)$, $(\bt,\gm)$. Such structure
constants can be defined in terms of Kronecker deltas, but the 
interesting case in $N\geq3$ dimensions is
\be
d^{\rho\tau\bt}_{\ka\nu\gm} = \eps^{\rho\tau\bt\mu_1..\mu_{N-3}}
\eps_{\ka\nu\gm\mu_1..\mu_{N-3}}
\ee
Then we can add an inhomogeneous term to the transformation law for the
$(4,2;1)$-type tensor field in (\ref{Rext}).
\bes
&&[L_\mu(m), R^{\lrst}_{\ka\nu}(n)] = (4,2;1)+
\nlb{LR}
&&\qquad+ d^{\rho\tau\al}_{\ka\nu\mu} m_\al m_\bt S_3^{\la\si\bt}(m+n)
+\symm(\la\rho,\si\tau),
\eens
where $\symm(\la\rho,\si\tau)$ stands for the three extra terms needed
to give the rhs the appropriate symmetries.
This follows immediately by specializing (\ref{MF}) to $\oj = gl(N)$
and (\ref{Tredef}). In three dimensions, 
\be
R^{\lrst}_{\ka\nu}(n)
= \eps^{\la\si\al} d^{\rho\tau\bt}_{\ka\nu\gm} \Gamma^\gm_{\al\bt}(n)
+ \symm(\la\rho,\si\tau),
\ee
where $\Gamma^\nu_{\si\tau}$ is the connection (\ref{lc-conn}).
The additional term in (\ref{LR}) is new. It can not be embedded in
a larger algebra using (\ref{addP}), because that would violate the
Jacobi identities \cite{Lar00}.

\subsection{ Dzhumadil'daev-Ovsienko-Roger cocycles in higher dimensions }
In this subsection I describe higher-dimensional generalizations of
the cocycles $\gamma_1$--$\gamma_3$ of (\ref{ORext}). 
Since such generalizations contain non-trivial extensions of $diff(1)$, 
these new cocycles are also non-trivial.
First tensor modules (\ref{tensor}) must be generalized to
translated tensor modules with dead indices. This concept is best
illustrated by an example. If
\bes
[L_\mu(m), \Phi^{\nu\rho}_{\si\tau}(n)]
&=& (n_\mu + \la m_\mu + r_\mu)\Phi^{\nu\rho}_{\si\tau}(m+n) +\nle
&&+ \dlt^\nu_\mu m_\la \Phi^{\la\rho}_{\si\tau}(m+n)
- m_\si \Phi^{\nu\rho}_{\mu\tau}(m+n),
\eens
we say that $\Phi^{\nu\rho}_{\si\tau}(m-r)$ is of type $(1,1;1)$ with
one dead upper index ($\rho$) and one dead lower index ($\tau$). The
remaining indices are, of course, alive.

1. Consider the redefinition
\bes
L_\mu(m) &\mapsto& L_\mu(m) + a A_\mu(m), \nle
[L_\mu(m), A_\nu(n)] &=& (n_\mu + (1-\la)m_\mu + r_\mu) A_\nu(m+n).
\eens
Thus, $A_\nu(m-r)$ is of type $(0,0;\la)$ with a dead lower index.
The limit $\la\to1$, $r_\mu=0$, $a(1-\la)=1$, gives rise to the cocycle
\bes
c(\xi,\eta) &=& A_\rho(\dmu\xmu\eta^\rho - \dnu\ynu\xi^\rho), \nle
c_{\mu\nu}(m,n) &=& m_\mu A_\nu(m+n) - n_\nu A_\mu(m+n),
\eens
which is a higher-dimensional generalization of $\gamma_1$.
The limit $r_\mu\to0$, $\la=1$, $ar_\mu = e_\mu$, yields
\bes
c(\xi,\eta) &=& A_\rho(e_\mu\xmu\eta^\rho - e_\nu\ynu\xi^\rho), \nle
c_{\mu\nu}(m,n) &=& e_\mu A_\nu(m+n) - e_\nu A_\mu(m+n),
\eens
This cocycle vanishes when $N=1$.

2. Consider the redefinition
$L_\mu(m) \mapsto L_\mu(m) + a m_\nu B^\nu_\mu(m)$,
where either $B^\nu_\mu(m-r)$ is of type $(1,0;\la)$ with a dead lower
index, or it is of type $(0,1;\la)$ with a dead upper index.
The limit $\la\to1$, $r_\mu=0$, $a(1-\la)=1$, gives rise to the cocycle
\bes
c(\xi,\eta) &=& B^\rho_\si(\dmu\xmu\drho\eta^\si - \dnu\ynu\drho\xi^\si),
\nle
c_{\mu\nu}(m,n) &=& m_\mu n_\rho B^\rho_\nu(m+n) 
- n_\nu m_\rho B^\rho_\mu(m+n).
\eens
This cocycle vanishes when $N=1$.
The limit $r_\mu\to0$, $\la=1$, $ar_\mu = e_\mu$, yields a trivial
cocycle.

3. Consider the redefinition
$L_\mu(m) \mapsto L_\mu(m) + a m_\rho m_\si K^{\rho\si}_\mu(m)$,
where $K^{\rho\si}_\mu(m-r)$ is of type $(2,1;\la)$, symmetric and
$\rho$ and $\si$, and all indices are alive.
As described above, the limit $\la\to1$, $r_\mu=0$, $a(1-\la)=1$, gives 
rise to the cocycles $\psi^W_1$ and $\psi^W_2$, for the trace and 
traceless parts, respectively. These are higher-dimensional 
generalizations of $\gamma_2$.
The limit $r_\mu\to0$, $\la=1$, $ar_\mu = e_\mu$, yields
\bes
c(\xi,\eta) &=& K^{\si\tau}_\rho(\dsi\dtau\xmu e_\nu \ynu -
 \dsi\dtau\ynu e_\mu\xmu), \nle
c_{\mu\nu}(m,n) &=& e_\nu m_\si m_\tau K^{\si\tau}_\mu(m+n)
 - e_\mu n_\si n_\tau K^{\si\tau}_\nu(m+n),
\eens
which is a higher-dimensional generalization of $\gamma_3$.
With $K^{\si\tau}_\rho = \dlt^{(\tau}_\rho S^{\si)}$, we obtain
\bes
c(\xi,\eta) &=& S^\si(\dsi\dmu\xmu e_\nu \ynu -
 \dsi\dnu\ynu e_\mu\xmu), \nle
c_{\mu\nu}(m,n) &=& (e_\nu m_\si m_\mu- e_\mu n_\si n_\nu) S^\si(m+n),
\eens
where $S^\rho$ is of type $(1,0;1)$. 

4. The natural way to generalize $\gamma_4$ and $\gamma_5$ would be to
redefine $L_\mu(m) \mapsto L_\mu(m) 
+ a m_\rho m_\si m_\tau D^{\rho\si\tau}_\mu(m)$,
where $D^{\rho\si\tau}_\mu(m-r)$ is of type $(3,1;\la)$ and totally
symmetric. However, as noted in (\ref{3redef}), this gives rise to a
trivial cocycle even when $\la=1$ and $r_\mu=0$, except in one dimension.
Hence I suspect that
$\gamma_4$ and $\gamma_5$ have no $N>1$ counterparts.

\subsection{ Anisotropic extensions}
\label{aniso}
In \cite{Lar97} I constructed two complicated cocycles satisfied by the
representations introduced by Eswara-Rao and Moody \cite{ERM94}.
It turned out \cite{Lar98,Lar99} that they could be obtained from the DRO 
algebra defined below (section \ref{DRO}), by imposing the 
second-class constraint
$L_f \approx 0$, $q^0(t) \approx t$. The former conditions can be
viewed as a first class constraint and the latter as a gauge condition.
Other cocycles can be found by replacing the gauge condition, as long
as the constraints together are second class, i.e. the Poisson
bracket matrix is invertible.

\section{ Extensions of $diff(N)\ltimes map(N,diff(d))$ }
\label{sec:MAP}
Replace $N$ by $N+d$ everywhere in the previous sections. 
The total space $\RR^{N+d}$ have coordinates $z^A = (x^\mu, y^i)$,
where greek indices $\mu,\nu,\rho,\si,\tau = 1, \ldots, N$ label 
horizontal (base space) directions, 
latin indices $i,j,k,\ell = 1, \ldots, d$ label vertical (target space)
directions, and capitals $A = (\mu,i)$, etc. label directions in total 
space. 
This induces splits $\d_A \equiv \d/\d z^A = (\dmu, \d_i) \equiv
(\d/\d x^\mu, \d/\d y^i)$, $\Xi^A(z) = (\xmu(x), X^i(x,y))$, 
$\LL_\Xi = (\Lxi, \J_X)$, etc. What makes this split a proper
embedding is that the horizontal components of the vector fields $\xmu(x)$
are taken to be independent of the vertical coordinates $y^i$, so
$\d_i\xmu = 0$.

An extension of $diff(N)\ltimes map(N,diff(d))$ has the form
\bes
[\Lxi,\Leta] &=& \LL_{[\xi,\eta]} + c(\xi,\eta), \nl
{[}\Lxi, \J_X] &=& \J_{\xmu\dmu X} + c(\xi,X), 
\nlb{LxX}
{[}\J_X, \J_Y] &=& \J_{[X,Y]} + c(X,Y).
\eens
Tensor densities are described by (\ref{tensor}), 
$\LL_\Xi = \Xi^A\d_A + \d_B\Xi^A T^B_A$, where $T^A_B$ satisfy $gl(N+d)$.
Hence
\bes
\Lxi &=& \xmu\dmu + \dnu\xmu T^\nu_\mu, 
\nlb{mtensor}
\J_X &=& X^i\d_i + \d_j X^i T^j_i + \dmu X^i T^\mu_i.
\eens
Since the $T^i_\nu$ component never enters any formulas, its value is
unimportant and may be set to zero. We can then perform a similarity
transformation $T^A_B \mapsto T^{\prime A}_B = \tilde S^A_C T^C_D S^D_B$,
with
\be
S^A_B = \begin{pmatrix} \dlt^\mu_\nu & 0 \\
 0 & \vareps\dlt^i_j \end{pmatrix} , \qquad
\tilde S^A_B = \begin{pmatrix} \dlt^\mu_\nu & 0 \\
0 & \vareps^{-1}\dlt^i_j \end{pmatrix},
\qquad
T^{\prime A}_B =\begin{pmatrix} T^\mu_\nu & \vareps T^\mu_j \\
 0 & T^i_j \end{pmatrix}.
\label{sim}
\ee
This amounts to multiplying the last term in (\ref{mtensor}) by
$\vareps$. For convenience, the transformation laws for a tensor field
of type $(1,1;1)$ in base space and $(1,1)$ in target space is given
explicitly; the general case follows readily.
\bes
[\Lxi, \Phi^{\si k}_{\tau \ell}(\phi_{\si k}^{\tau \ell})]
&=& \Phi^{\si k}_{\tau \ell}( \xmu\dmu\phi_{\si k}^{\tau \ell} 
+ \dsi\xmu\phi_{\mu k}^{\tau \ell} 
- \dnu\xi^\tau \phi_{\si k}^{\nu \ell}), 
\label{basetens} \\
{[}\J_X, \Phi^{\si k}_{\tau \ell}(\phi_{\si k}^{\tau \ell})]
&=& \Phi^{\si k}_{\tau \ell}( X^i\d_i\phi_{\si k}^{\tau \ell} 
+ \d_k X^i\phi_{\si i}^{\tau \ell} 
- \d_j X^\ell \phi_{\si k}^{\tau j} +
\nlb{targtens}
&&+ \vareps \dsi X^i\phi_{i k}^{\tau \ell}),
\eens
where $\phi_{\si k}^{\tau \ell}(x,y)$ is an arbitrary function on
total space.
Of course, a similarity transformation does not bring anything essentially
new, but we can set $\vareps=0$ in (\ref{targtens}), corresponding
to a singular matrix $S^A_B$. Base space and target space indices then
decouple which makes the transformation laws particularly simple.

The similarity transformation amounts to a rescaling of $\dsi X^i$ by
$\vareps$ without affecting other components of $\d_B \Xi^A$. This is
equivalent to rescaling $X^i$ by $\vareps$ and $\d_j$ by $\vareps^{-1}$.
$\d_j \xmu$ would also rescale by $\vareps^{-1}$, but this is no problem
since it vanishes anyway. What is a problem is that $\d_j \d_k X^i$
also rescales by $\vareps^{-1}$. Since all cocycles contain such terms,
we can in fact not put $\vareps=0$, but it will become possible in
the next section. For the remainder of this section, we set $\vareps=1$.

The restrictions of the generic extensions are as follows.

\smallskip\noindent\underline{$\bar\psi^W_3$}:
\bes
c(\xi,X) &=& S^\rho(\drho\dmu\xmu \d_iX^i), \nle
c(X,Y) &=& S^\rho(\drho\d_iX^i\d_jY^j) + S^k(\d_k \d_iX^i\d_jY^j).
\eens

\smallskip\noindent\underline{$\bar\psi^W_4$}:
\bes
c(\xi,X) &=& 0, \nle
c(X,Y) &=& S^\rho(\drho\d_jX^i\d_iY^j) + S^k(\d_k \d_jX^i\d_iY^j).
\eens
Here, $S^C(\phi_C) = S^\rho(\frho)+ S^k(\phi_k)$ is a tensor density
in total space of type $(1,0;1)$, satisfying the auxiliary condition
$S^C(\d_C\phi) \equiv 0$. Explicitly, the transformation laws
are given by
\bes
[\Lxi, S^\rho(\frho)] &=& S^\rho(\xmu\dmu\frho + \drho\xmu\phi_\mu), \nl
{[}\Lxi, S^k(\phi_k)] &=& S^k(\xmu\dmu\phi_k), \nle
{[}\J_X, S^\rho(\frho)] &=& S^\rho(X^i\d_i\frho), \nl
{[}\J_X, S^k(\phi_k)] &=& S^k(X^i\d_i\phi_k + \d_k X^i\phi_i)
 + S^\rho(\drho X^i\phi_i),
\eens
where $S^\rho(\drho\phi) + S^k(\d_k\phi) \equiv 0$.

\smallskip\noindent\underline{$\psi^W_1$}:
\bes
c(\xi,X) &=& S^\rho(\drho\dmu\xmu \d_iX^i - \dmu\xmu \drho\d_iX^i)
- S^k(\dmu\xmu \d_k\d_iX^i), \nl
c(X,Y) &=& S^\rho(\drho\d_iX^i\d_jY^j - \d_iX^i\drho\d_jY^j) \\
&&+ S^k(\d_k \d_iX^i\d_jY^j - \d_iX^i\d_k\d_jY^j).
\eens
where $S^C$ is as above but the closedness condition is no longer
necessary.

\smallskip\noindent\underline{$\psi^W_2$}:
\bes
c(\xi,X) &=& K^{(AB)}_i(\dmu\xmu\d_A\d_B X^i)
 - K^{(\si\tau)}_\mu(\dsi\dtau\xmu\d_iX^i), \nle
c(X,Y) &=& K^{(AB)}_j(\d_iX^i\d_A\d_BY^j)
 - K^{(AB)}_i(\d_A\d_BX^i\d_jY^j),
\eens
where $K^{(AB)}_C$ is a tensor field of type $(2,1;1)$.

\smallskip\noindent\underline{$\psi^W_3$--$\psi^W_{10}$}:
\bes
c(\xi,X) &=& 
R^{(\la\rho)(CD)}_{\mu i}(\dla\drho\xmu\d_C\d_D X^i), \nle
c(X,Y) &=&
R^{(AB)(CD)}_{ij}(\d_A\d_BX^i\d_C\d_DY^j), 
\eens
where $R^{(AB)(CD)}_{EF}$ is a tensor field of type $(4,2;1)$.

\section{ Extensions of $diff(N)\ltimes map(N,gl(d))$ }
Now consider the subalgebra $gl(d)\subset diff(d)$, with
vertical vector fields $X = X^i(x,y)\d_i = X^i_j(x) y^j\d_i$.
In the previous section, we substitute $\d_jX^i = X^i_j$,
$\d_j\d_k X^i = 0$. The algebra formally takes the same form
(\ref{LxX}), but now $[X,Y] = (X^i_kY^k_j - X^k_jY^i_k)y^j\d_i$.
Tensor fields are decomposed into components which are homogeneous
in $y^i$, e.g.,
\be
\Phi^{\si k}_{\tau \ell}(\phi_{\si k}^{\tau \ell}) =
\sum_{n=0}^\infty
\Phi^{\si k|\mn}_{\tau \ell}(\phi_{\si k|\mn}^{\tau \ell}),
\ee
where $\phi_{\si k|\mn}^{\tau \ell}(x)$ is a function independent of
the vertical coordinate $y^i$ and
\be
\Phi^{\si k|\mn}_{\tau \ell}(\cdot) \equiv
\Phi^{\si k}_{\tau \ell}(y^{m_1}\ldots y^{m_n}\,\cdot).
\ee
The base space transformation law (\ref{basetens}) is unchanged, whereas
(\ref{targtens}) is replaced by
\bes
&&[\J_X, \Phi^{\si k|\mn}_{\tau \ell}(\phi_{\si k|\mn}^{\tau \ell})]
= \Phi^{\si k|\mn}_{\tau \ell}( X^i_k\phi_{\si i|\mn}^{\tau \ell}-\\
\bl- X^\ell_j \phi_{\si k|\mn}^{\tau j} 
+\sum_{r=1}^n X^i_{m_r} \phi_{\si k|m_1..i..m_n}^{\tau \ell}) 
+ \vareps \Phi^{\si k|\mn j}_{\tau \ell}
 (\dsi X^i_j\phi_{i k|\mn}^{\tau \ell}).
\eens
In this section we can set $\vareps=0$, since the dangerous term
$\d_j\d_k X^i = 0$ anyway. Because $X^i_j = \d_j X^i$, this amounts to
rescalings of $y^i$ by $\vareps$ and of $\d_j$ by $\vareps^{-1}$, and 
hence $\Phi^{\si k|\mn}_{\tau \ell}(\cdot)$ must be multiplied by 
$\vareps^n$.

With any value of $\vareps$, transformation laws are readily read off
from the index structure. In particular, target space indices transform
in the same way independent of if they appear to the left or to the right 
of a vertical bar.

\smallskip\noindent\underline{$\bar\psi^W_3$}:
\bes
c(\xi,X) &=& S^\rho(\drho\dmu\xmu X^i_i), \nle
c(X,Y) &=& S^\rho(\drho X^i_iY^j_j).
\eens

\smallskip\noindent\underline{$\bar\psi^W_4$}:
\bes
c(\xi,X) &=& 0, \nle
c(X,Y) &=& S^\rho(\drho X^i_jY^j_i).
\eens
Note that only the horizontal component of the one-form $S^C(\phi_C)$
appears, and that its argument is independent of $y^i$. Therefore,
we can limit our attention to $S^\rho(\frho)$, where $\frho(x)$ is
independent of the vertical coordinates and $\phi_i = 0$.
The transformation laws read
\bes
[\Lxi, S^\rho(\frho)] &=& S^\rho(\xmu\dmu\frho + \drho\xmu\phi_\mu), \nle
{[}\J_X, S^\rho(\frho)] &=& 0,
\eens
where $S^\rho(\drho\phi) \equiv 0$.

\smallskip\noindent\underline{$\psi^W_1$}:
\bes
c(\xi,X) &=& S^\rho(\drho\dmu\xmu X^i_i - \dmu\xmu \drho X^i_i), \nle
c(X,Y) &=& S^\rho(\drho X^i_iY^j_j - X^i_i\drho Y^j_j).
\eens
where $S^\rho$ is as above but the closedness condition is no longer
necessary.

\smallskip\noindent\underline{$\psi^W_2$}:
\bes
c(\xi,X) &=& \vareps K^{(\si\tau)|j}_i(\dmu\xmu\dsi\dtau X^i_j) 
+ 2K^{(\si j)}_i(\dmu\xmu\dsi X^i_j) -\nl
&&- K^{(\si\tau)}_\mu(\dsi\dtau\xmu X^i_i), \\
c(X,Y) &=& \vareps K^{(\si\tau)|k}_j(X^i_i\dsi\dtau Y^j_k)
+ 2K^{(\si k)}_j(X^i_i\dsi Y^j_k) - \XxY,
\eens
where $K^{(\si\tau)|j}_i(\cdot) = K^{(\si\tau)}_i(y^j \,\cdot)$.
The two cocycles that survive when $\vareps=0$ are independent.

\smallskip\noindent\underline{$\psi^W_3$--$\psi^W_{10}$}:
\bes
c(\xi,X) &=& 
\vareps R^{\lrst|j}_{\mu i}(\dla\drho\xmu\dsi\dtau X^i_j) +
2R^{(\la\rho)(\si j)}_{\mu i}(\dla\drho\xmu\dsi X^i_j), \nl
c(X,Y) &=&
\vareps^2 R^{\lrst|k\ell}_{ij}(\dla\drho X^i_k\dsi\dtau Y^j_\ell) +
\label{gl310}\\
&&+ 2\vareps (
 R^{(k\rho)(\si\tau)|\ell}_{ij}(\drho X^i_k\dsi\dtau Y^j_\ell) +
 R^{(\la\rho)(\ell\tau)|k}_{ij}(\dla\drho X^i_k\dtau Y^j_\ell) ) +\nl
&&+ 4R^{(k\rho)(\ell\tau)}_{ij}(\drho X^i_k\dtau Y^j_\ell),
\eens
where
\bes
R^{\lrst|j}_{\mu i}(\cdot) &=& 
R^{\lrst}_{\mu i}(y^j \,\cdot), \nl
R^{\lrst|k\ell}_{ij}(\cdot) &=&
R^{\lrst}_{\mu i}(y^ky^\ell \,\cdot), \nle
R^{(k\rho)(\si\tau)|\ell}_{ij}(\cdot) &=&
R^{(k\rho)(\si\tau)}_{ij}(y^\ell \,\cdot).
\eens
The two cocycles that survive when $\vareps=0$ are independent, and the
the last term in (\ref{gl310}) is recognized as the MF cocycle (\ref{MF})
for $map(N,gl(d))$.

\section{ Extensions of $diff(N)\ltimes map(N,\oj)$ }
\label{sec:map}
Assume that the finite-dimensional Lie algebra $\oj$ has a 
$d$-dimensional representation with matrices $\si^a = (\si^{ia}_j)$.
In the previous section, we substitute $X^i_j = X_a \si^{ia}_j$.
Set $\tr \si^a = \si^{ia}_i = z_M \dlt^a$, where either 
$\dlt^c f^{ab}{}_c = 0$ or $z_M=0$, 
and $\tr \si^a\si^b = \si^{ia}_j\si^{jb}_i = y_M \dlt^{ab}$.
Now $[X,Y]_c = if^{ab}{}_cX_aY_b$, $X^i_i = z_M\dlt^aX_a$ and 
$X^i_jY^j_i = y_M \dlt^{ab}X_aY_b$.
Tensor fields are given by
\bes
\Lxi &=& \xmu\dmu + \dnu\xmu T^\nu_\mu, 
\nle
\J_X &=& X_a \si^{ia}_j y^j\d_i +  X_a \si^{ia}_j T^j_i 
 + \vareps\dmu X_a \si^{ia}_j T^\mu_i.
\eens

\smallskip\noindent\underline{$\bar\psi^W_3$}:
\bes
c(\xi,X) &=& z_M\dlt^a S^\rho(\drho\dmu\xmu X_a), \nle
c(X,Y) &=& z^2_M\dlt^a\dlt^b S^\rho(\drho X_aY_b).
\eens
\smallskip\noindent\underline{$\bar\psi^W_4$}:
\bes
c(\xi,X) &=& 0, \nle
c(X,Y) &=& y_M \dlt^{ab} S^\rho(\drho X_aY_b).
\eens
In particular, in one dimension we get
\bes
[L_m, J^a_n] &=& nJ^a_{m+n} + z_M\dlt^a m^2 \dlt_{m+n}, \nle
{[}J^a_m, J^b_n] &=& if^{ab}{}_c J^c_{m+n} 
+ z^2_M\dlt^a\dlt^b m\dlt_{m+n} + y_M \dlt^{ab} m \dlt_{m+n}.
\eens
The last term is recognized as the Kac-Moody cocycle. The other two
are not so well known, because they vanish for $\oj$ semisimple.
However, all three cocycles are non-trivial.

\smallskip\noindent\underline{$\psi^W_1$}:
\bes
c(\xi,X) &=& z_M\dlt^a S^\rho(\drho\dmu\xmu X_a - \dmu\xmu \drho X_a), 
\nle
c(X,Y) &=& z^2_M\dlt^a\dlt^b S^\rho(\drho X_aY_b - X_a\drho Y_b).
\eens

\smallskip\noindent\underline{$\psi^W_2$}:
\bes
c(\xi,X) &=& \vareps \si^{ia}_j K^{(\si\tau)|j}_i(\dmu\xmu\dsi\dtau X_a) 
+ 2\si^{ia}_j K^{(\si j)}_i(\dmu\xmu\dsi X_a) -\nl
&&- z_M\dlt^a K^{(\si\tau)}_\mu(\dsi\dtau\xmu X_a), \\
c(X,Y) &=& \vareps z_M \dlt^a \si^{jb}_k 
 K^{(\si\tau)|k}_j(X_a\dsi\dtau Y_b) +\nl
&&+ 2\dlt^a \si^{jb}_k K^{(\si k)}_j(X_a\dsi Y_b) - \XxY.
\eens
The two cocycles that survive when $\vareps=0$ are independent.

\smallskip\noindent\underline{$\psi^W_3$--$\psi^W_{10}$}:
\bes
c(\xi,X) &=& 
\vareps \si^{ia}_j R^{\lrst|j}_{\mu i}(\dla\drho\xmu\dsi\dtau X_a) +
2\si^{ia}_jR^{(\la\rho)(\si j)}_{\mu i}(\dla\drho\xmu\dsi X_a), \nl
c(X,Y) &=&
\vareps^2 \si^{ia}_k\si^{jb}_\ell R^{\lrst|k\ell}_{ij}
 (\dla\drho X_a\dsi\dtau Y_b) +\nl
&&+ 2\vareps \si^{ia}_k\si^{jb}_\ell 
 (R^{(k\rho)(\si\tau)|\ell}_{ij}(\drho X_a\dsi\dtau Y_b) +\\
&&\qquad+ R^{(\la\rho)(\ell\tau)|k}_{ij}(\dla\drho X_a\dtau Y_b) )+\nl
&&+ 4\si^{ia}_k\si^{jb}_\ell 
 R^{(k\rho)(\ell\tau)}_{ij}(\drho X_a\dtau Y_b).
\eens
The two cocycles that survive when $\vareps=0$ are independent, and the
latter is recognized as the MF cocycle (\ref{MF}) for $map(N,\oj)$.

\section{Extensions of $diff(N)\oplus diff(1)$: DRO algebra}
\label{DRO}
The DRO (Diffeomorphism, Reparametrization, Observer) algebra $DRO(N)$ 
was introduced in \cite{Lar98,Lar99} as an extension of 
$diff(N)\oplus diff(1)$ by the observer's trajectory $q^\mu(t)$. 
The reason for giving this algebra a special name is its importance 
for Fock
representations of $diff(N)$. Expand all fields in a Taylor series around
$q^\mu(t)$, where $t\in S^1$. The Taylor coefficients, or {\em jets}, are
\be
\Phi_{,\mm}(t) = \d_\mm\Phi(q(t)) \equiv 
\d_1^{m_1} \ldots \d_N^{m_N} \Phi(q(t)).
\label{jet}
\ee
where $\mm = (m_1,\ldots,m_N)$ is a multi-index.
Note that the jets depend on $t$ although
the field $\Phi(x)$ does not, since this dependence enters through 
the expansion point. 
In \cite{Lar98} I took the space of $p$-jets $\Phi_{,\mm}(t)$,
with $|\mm| = \sum_{\mu=1}^N m_\mu \leq p$, as the starting point for
the Fock construction. This leads to consistent results because the jet
space consists of finitely many functions of a single variable $t$.
The full DRO algebra acts naturally on the jets;
the additional $diff(1)$ factor describes
reparametrizations of the observer's trajectory.

Any extension of $diff(N)\oplus diff(1)$ has the form
\bes
[\Lxi,\Leta] &=& \LL_{[\xi,\eta]} + c(\xi,\eta), \nl
{[}\Lxi, L_f] &=& c(\xi,f), 
\nlb{Lxf}
{[}L_f, L_g] &=& L_{[f,g]} + c(f,g),
\eens
where $f = f(t)d/dt$ is a vector field on the circle and 
$[f,g] = (f\dot g - g\dot f)d/dt$.

We embed $diff(N)\oplus diff(1)\subset diff(N+1)$ 
in the natural way: set $z^A \equiv (z^\mu, z^0) = (x^\mu, t)$, 
$\d_A = (\dmu, d/dt)$, $\Xi^A(z) = (\xmu(x), f(t))$, 
$\LL_\Xi = (\Lxi, L_f)$.
Tensor densities restrict to
\bes
\Lxi &=& \xmu\dmu + \dnu\xmu T^\nu_\mu, \nle
L_f &=& f{d\/dt} + \dot f T^0_0,
\eens
where $T^0_0$ was called the {\em causal weight} in 
\cite{Lar97}--\cite{Lar99}. Thus, both the $T^0_\mu$ and $T^\mu_0$
components of the $gl(N+1)$ generator $T^A_B$ decouple.

\smallskip\noindent\underline{$\bar\psi^W_3$}:
\bes
c(\xi,f) &=& S^\rho(\drho\dmu\xmu \dot f), 
\nlb{DRO3}
c(f,g) &=& S^0(\ddot f\dot g).
\eens

\smallskip\noindent\underline{$\bar\psi^W_4$}:
\bes
c(\xi,f) &=& 0, 
\nlb{DRO4}
c(f,g) &=& S^0(\ddot f\dot g).
\eens
With $\phi(x)$ independent of $t$ and $f(t)$ independent of $x^\mu$,
closedness implies
\be
S^\rho(\drho\phi f) + S^0(\phi\dot f) \equiv 0.
\label{DROclosed}
\ee
In particular, $S^0(\dot f) \equiv 0$, so $S^0(f) \propto \int dt\ f(t)$ 
and $c(f,g)$ is the Virasoro cocycle in both cases.

Thus $DRO(N)$ has four independent Virasoro-like cocycles, namely the
terms proportional to $c_1$, $c_2$, $c_3$ and $c_4$ in the notation of 
\cite{Lar98}.
In the notation of the present paper, $c_1 = \psi^W_4$, $c_2 = \psi^W_3$,
$c_3 = c(\xi,f)$ from (\ref{DRO3}),
and $c_4 = c(f,g)$ from (\ref{DRO3}) or (\ref{DRO4}). As described
in subsection \ref{aniso}, we can eliminate reparametrizations
by a second class constraint, trading the last two cocycles for 
anisotropic cocycles of $diff(N)$.

\smallskip\noindent\underline{$\psi^W_1$}:
\bes
c(\xi,f) &=& S^\rho(\drho\dmu\xmu \dot f)- S^0(\dmu\xmu \ddot f), \nl
c(f,g) &=& S^0(\ddot f\dot g - \dot f\ddot g).
\eens
where (\ref{DROclosed}) no longer holds. The second formula is the
Virasoro generalization (\ref{Svir}).

\smallskip\noindent\underline{$\psi^W_2$}:
\bes
c(\xi,f) &=& K^{00}_0(\dmu\xmu\ddot f) 
 - K^{(\si\tau)}_\mu(\dsi\dtau\xmu\dot f), \nle
c(f,g) &=& K^{00}_0(\dot f\ddot g - \ddot f\dot g).
\eens

\smallskip\noindent\underline{$\psi^W_3$--$\psi^W_{10}$}:
\bes
c(\xi,f) &=& R^{(\la\rho)}_\mu(\dla\drho\xmu\ddot f), \nle
c(f,g) &=& 0.
\eens
where $R^{(\la\rho)}_\mu$ is a tensor field of type $(3,1;1)$.

\section{Conclusion}
In this paper I have reviewed Dzhumadil'daev's exhaustive classification 
of tensor extensions of $diff(N)$ and subalgebras, 
extended it beyond tensor modules, and studied the
chain of restrictions down to $map(N,\oj)$. The method proves 
existence for the cocycles of the subalgebras, but it neither proves
non-triviality nor exhaustion. However, since the extensions obtained
in the last step are in fact recognized as non-trivial (Kac-Moody,
MF, etc.), the entire chain is non-trivial. Moreover,
I tautologically exhaust the class of subalgebra cocycles with values in
tensor modules, which
can be lifted to the diffeomorphism algebra in total space.

The construction of projective Fock modules
of $diff(N)$ was initiated in \cite{ERM94} and further developped in
\cite{Lar97}--\cite{Lar99}. By restriction, this gives Fock modules
of subalgebras, of the type described in 
\cite{EMY92,MEY90} and in the papers just cited. 
Berman and Billig \cite{BB98} constructed another type of module, 
postulating the two cocycles $\bar\psi^W_3$ and $\bar\psi^W_4$ from the
outset. It seems likely that a deep generalization of their modules
exists, if one starts from the four inequivelent Virasoro-like
extensions of $DRO(N)$ instead. The $diff(1)$ factor should then 
provide the necessary extra-grading.
Finally, Fock modules for extensions 
of current algebras that are similar to, but different from, the 
Mickelsson-Faddeev algebra have recently been constructed \cite{Lar00}.

This work can be extended in several directions. One can consider
subalgebras of $diff(N)$ such as algebras of divergence-free, 
Hamiltonian or contact vector fields, or superize by letting some
coordinates become fermionic. I expect no essential difficulties here, 
except that I am not aware of any classification of 
extensions of superdiffeomorphism algebras.

\section*{Acknowledgments}
I am grateful A. Dzhumadil'daev for explaining his results to me, in
particular the special two-dimensional cocycles.

\end{document}